\documentclass[a4paper,11pt]{article}
\pdfoutput=1 

\usepackage{jheppub} 
\usepackage{amsmath}
\usepackage[T1]{fontenc} 
\usepackage{booktabs}

\title{Study of Electroweak Phase Transition in Exotic Higgs Decays at the CEPC}

\author[a,b,g]{Zhen Wang,}
\author[a,b]{Xuliang Zhu,}
\author[f]{Elham E Khoda,}
\author[f]{Shih-Chieh Hsu,}
\author[h]{Nikolaos Konstantinidis}
\author[f]{Ke Li,}
\author[a,b,e,i]{Shu Li,}
\author[a,b,c,d]{Michael J. Ramsey-Musolf,}
\author[a,b]{Yanda Wu,}
\author[h]{Yuwen E. Zhang}

\affiliation[a]{ Tsung-Dao Lee Institute,
Shanghai Jiao Tong University,\\520 Shengrong Road, Shanghai 201210, China}
\affiliation[b]{ Institute of Nuclear and Particle Physics, School of Physics and Astronomy,\\Key Laboratory for Particle Astrophysics and Cosmology (MOE), Shanghai Key Laboratory for Particle Physics and Cosmology (SKLPPC), Shanghai Jiao Tong University,\\800 Dongchuan Road, Shanghai 200240, China}
\affiliation[c]{ Amherst Center for Fundamental Interactions,
Department of Physics, University of Massachusetts,\\Amherst, MA 01003, USA}
\affiliation[d]{ Kellogg Radiation Laboratory,
California Institute of Technology,\\Pasadena, CA 91125 USA}
\affiliation[e]{ Center for High Energy Physics, Peking University,\\5 Yiheyuan Road, Beijing 100871, China}
\affiliation[f]{ Department of Physics, University of Washington,\\Seattle 98195-1560, USA}
\affiliation[g]{ Department of Physics, Duke University,\\Durham, North Carolina 27708, USA}
\affiliation[h]{ Department of Physics and Astronomy, University College London,\\Gower Street, London, WC1E 6BT, UK}
\affiliation[i]{ School of Mechanical and Electronic Engineering, Suzhou University,\\Suzhou 234000, Anhui, China}

\emailAdd{elham.e.khoda@cern.ch}
\emailAdd{schsu@uw.edu}
\emailAdd{shuli@sjtu.edu.cn}

\newcommand{\MGMCatNLO}{\textsc{MadGraph5}\_aMC@NLO}
\newcommand{\whizard}{\textsc{Whizard}}
\newcommand{\pythia}{\textsc{Pythia8}}
\newcommand{\feynrules}{\textsc{FeynRules}}

\abstract{
A strong first-order electroweak phase transition (EWPT) can be induced by light new physics weakly coupled to the Higgs. This study focuses on a scenario in which the first-order EWPT is driven by a light scalar $s$ with a mass between 15-60 GeV. A search for exotic decays of the Higgs boson into a pair of spin-zero particles, $h \to ss$, where the $s$-boson decays into $b$-quarks promptly is presented. The search is performed in events where the Higgs boson is produced in association with a $Z$ boson, giving rise to a signature of two charged leptons (electrons or muons) and multiple jets from $b$-quark decays. The analysis is considering a scenario of analysing 5000 fb$^{-1}$  $e^+ e^-$ collision data at $\sqrt{s} = 240 $ GeV from the Circular Electron Positron Collider (CEPC). This study with $4b$ final state conclusively tests the expected sensitivity of probing the light scalars in the CEPC experiment. The sensitivity reach is significantly larger than that can be achieved at the LHC.}

\notoc

\makeatletter
\gdef\@fpheader{Submitted to the  Proceedings of the US Community Study\\
on the Future of Particle Physics (Snowmass 2021)}
\makeatother

\begin{document} 
\maketitle
\flushbottom

\section{Introduction}
\label{sec:theory}
One of the open questions at the particle physics and cosmology frontier is to determine the thermal history of ElectroWeak Symmetry Breaking (EWSB)  in the early universe. In the Standard Model (SM), the lattice simulations indicate that the EWSB occurred via a cross-over transition \cite{KAJANTIE1996189,KAJANTIE1997413,PhysRevLett.77.2887,PhysRevLett.82.21}, which cannot provide the necessary conditions for electroweak baryogenesis \cite{Morrissey_2012} or a source for a stochastic gravitational wave background. The latter could be observed at future space-based experiment facilities such as LISA \cite{Caprini_2016}. It is interesting to ask how the new particles beyond the SM (BSM) can change this electroweak thermal history. A strong first order electroweak phase transition (SFOEWPT) is expected within BSM scenarios, thereby  supplying one of the necessary conditions for electroweak baryogenesis and gravitational wave background \cite{Morrissey_2012, Caprini_2016}. Any new particles involved in such alternative thermal history cannot have  masses too heavy with respect to the electroweak temperature scale, nor can they interact too weakly with the SM Higgs boson \cite{Ramsey-Musolf:2019lsf}. In this work, we consider a lighter BSM singlet particle $s$ coupled to the SM Higgs $h$ to catalyze a SFOEWPT. Specifically, we consider this new scalar's mass to be lighter than half of SM Higgs boson, thereby enabling exotic Higgs decay modes. As shown in Ref.\cite{kosaczuk_2020}, there exists a lower bound on the corresponding exotic Higgs decay branching ratio as a function of new scalar mass. See Ref. \cite{Carena:2019une} for a complementary study of such exotic Higgs decays as a signature of a SFOEWPT. 

Many high energy experimental facilities, such as the LHC and potential future Higgs-factory-like lepton colliders, could search for the presence of such light new scalar particles \cite{kosaczuk_2020,Carena:2019une}.
Moreover, future $e^{+}e^{-}$ colliders should have sufficient sensitivity to probe the SFOEWPT for new scalar masses down to at least $\sim$10 GeV \cite{kosaczuk_2020,Carena:2019une,2020_1911}. Under such motivation, we propose to perform a detailed study of a future exotic Higgs decay search with the reference detector simulation of the Circular Electron Positron Collider (CEPC).

We are interested in the strong first-order electroweak phase transition with the light real singlet, $S$, coupled with the SM Higgs, $H$. The relevant scalar potential involving $S$ and $H$ reads \cite{Profumo_2007,kosaczuk_2020}:

\begin{equation}
    V=-\mu^{2}|H|^{2}+\lambda|H|^{4}+\frac{1}{2} a_{1}|H|^{2} S+\frac{1}{2} a_{2}|H|^{2} S^{2}+b_{1} S+\frac{1}{2} b_{2} S^{2}+\frac{1}{3} b_{3} S^{3}+\frac{1}{4} b_{4} S^{4}
\end{equation}

After electroweak symmetry breaking, the fields can be parametrized as

\begin{equation}       
H=\frac{1}{\sqrt{2}}\left(                 
  \begin{array}{c}   
    0 \\  
    v+h \\  
  \end{array}
\right), \quad S=v_s+s  \ \ \ ,               
\end{equation}
where $v=246$ GeV is the vacuum expectation value (VEV) for the Higgs field at zero temperature and $v_s$ is the VEV for the singlet. As we introduce the tadpole term in the potential, we set $v_s=0$.

The two scalar fields $h$ and $s$ can mix to produce the mass eigenstates
\begin{eqnarray}
h_{1} &=h \cos \theta+s \sin \theta \nonumber \\
h_{2} &=-h \sin \theta+s \cos \theta ,
\end{eqnarray}

where $h_1$ and $h_2$ are separately singlet-like and SM-like Higgs particle, with the masses $m_1$ and $m_2$, respectively. We consider the case in which the SM-like Higgs $h_2$ decays visibly and promptly. This requires the mixing angle of the SM Higgs and the light scalar $\cos{\theta} \neq 0$. We choose $\cos{\theta} = 0.01$, since as long as the magnitude is small, the EWPT region with successful tunneling is insensitive to the precise value, and this value is large enough for $h_1$ to decay promptly \cite{kosaczuk_2020}. 

The total width of the singlet-like scalar and SM-like Higgs are

\begin{equation}
\begin{aligned}
    &\Gamma(h_1)=\cos^2{\theta}\Gamma(h_{2}) \\
    &\Gamma\left(h_{2}\right)=\left.\sin ^{2} \theta \Gamma^{\mathrm{SM}}\right|_{m_{2}}+\Gamma\left(h_{2} \rightarrow h_{1} h_{1}\right) \ \ \ ,
\end{aligned}
\end{equation}
where the exotic decay partial width is
\begin{equation}
    \Gamma\left(h_{2} \rightarrow h_{1} h_{1}\right)=\frac{1}{32 \pi^{2} m_{2}} \lambda_{211}^{2} \sqrt{1-\frac{4 m_{1}^{2}}{m_{2}^{2}}}
\end{equation}

The $h_2\to h_1 h_1$ decay is governed by the cubic coupling

\begin{equation}
    \lambda_{211}=2 s^{2} c b_{3}+\frac{a_{1}}{2} c\left(c^{2}-2 s^{2}\right)+\left(2 c^{2}-s^{2}\right) s v a_{2}-6 \lambda s c^{2} v \ \ \ ,
\end{equation}
where $s\equiv\sin\theta$ and $c\equiv\cos\theta$. As shown in Ref \cite{kosaczuk_2020}, in this singlet extension model, only $a_2, b_3, b_4$ in potential $V$ are free parameters; the other  parameters can be derived from this three parameters. The parameter region that satisfy a strong first order electroweak phase transition for this model have been worked out \cite{kosaczuk_2020}, whose results we have reproduced for this study.

\begin{figure}
    \centering
    \includegraphics[scale=0.6]{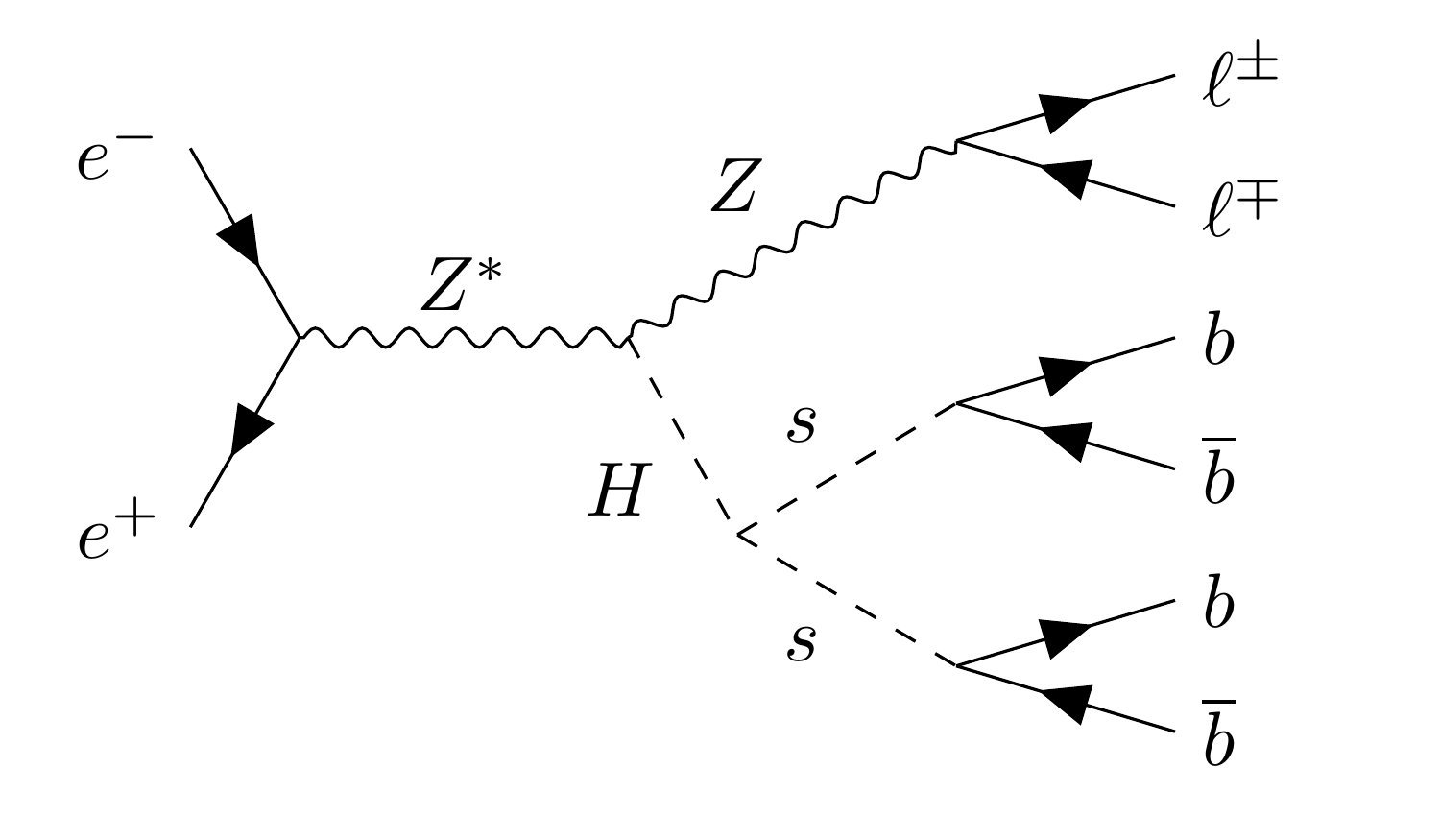}
    \caption{Representative tree-level Feynman diagram for the $ZH$ production process with subsequent decays $Z \to \ell \ell~(\ell = e,\mu)$ and $H \to s s \to 4b$.}
    \label{fig:sig_feynman}
\end{figure}

Under such motivation, we propose to perform a detailed study of a future exotic Higgs decay search with the reference detector simulation of CEPC, which is presently one of the major future candidate lepton collider experiments providing rich
$e^{+}e^{-}$ collision data at a center-of-mass energy of 240 GeV. A full consideration of the current CEPC detector geometry and condition effects are taken into account. A study of the search sensitivity of the Higgs exotic decays in $4b$ final states is carried out. At the center-of-mass energy of 240 GeV, the dominant Higgs production channel is the $ZH$ process, where the SM Higgs boson is produced in association with a $Z$ boson. Only the exotics decay mode of the SM Higgs boson ($H$) into a pair of singlet-like exotics Higgs ($s$) is considered. The search focuses on $ZH$ processes with $Z \to \ell \ell ~(\ell = e,\mu)$ and $H \to ss \to 4b$ as shown in Fig. \ref{fig:sig_feynman}. The $4b$ final state is chosen for this study since it has the highest branching ratio; so will be easier to probe in the clean environment of CEPC. Such study will help to further strengthen the BSM aspect of the physics motivation for CEPC in addition to its important role of SM Higgs factory for precision physics.

\section{Detector description}
\label{sec:detector}

The CEPC detector concept is designed to meet the requirements by a various of physics programs especially for precision measurements. From innermost to outermost, it consists of a low material tracking system, a high granularity calorimeter system surrounded by a 3 Tesla superconducting solenoid and a flux return yoke embedded with a muon detector. The tracking system is made by a silicon pixel vertex detector, a silicon inner tracker, a Time Projection Chamber (TPC) and a silicon external tracker. The charged track can be reconstructed precisely and the momentum resolution is $\Delta(1/p_T)~2\times10^5$ GeV$^{-1}$. Moreover the ionization energy loss ($dE/dx$) can be measured to provide good particle identification efficiency and purity. The calorimeter system is made by a silicon-tungsten sampling Electromagnetic Calorimeter and a steel-Glass Resistive Plate Chamber (GRPC) sampling Hadronic Calorimeter. It can provide a resolution of 3-5\% for jet energies between 20 and 100GeV. Further more it can provide a 50 ps resolution for time measurement and can be used for particle identification. The muon detector is made by RPC with an iron yoke serving as the magnetic flux return.

\section{Event simulation and object reconstruction}
\label{sec:MC}

\MGMCatNLO~\cite{Alwall:2014hca} and \whizard~\cite{Kilian:2007gr} Monte Carlo event generators are used to simulate the signal and background events. The singlet-like exotics Higgs signal model described in Sec. \ref{sec:theory} is implemented and import to \MGMCatNLO using \feynrules~ \cite{Alloul:2014}. The signal events are generated with \MGMCatNLO at leading order (LO), and the parton showering and hadronization modelling is done using \pythia~\cite{Sjostrand:2014zea}. Signal samples are generated for different single-like exotics Higgs mass starting from 15 GeV to 60 GeV in steps of 5 GeV. The SM background events are generated using \whizard, and the parton showering and hadronization modelling is also done using \pythia. All background and signal samples are generated at non-polarized electron-positron collision at $\sqrt{s} = 240 ~\mathrm{GeV}$. Samples simulated by CEPC are used to model all the two fermions, four fermions and the SM Higgs backgrounds. The detector simulation is performed by Mokka~\cite{MoradeFreitas:2002kj}, a Geant4~\cite{GEANT4:2002zbu} based detector simulator. The simulated hits are digitized and reconstructed with ArborPFA~\cite{Ruan:2018yrh}.

The charged leptons, like electron and muon, are identified by the Lepton Identification in Calorimeter with High Granularity~(LICH) algorithm ~\cite{Yu:2017mpx} which is designed for the Higgs factory. 
The overall identification efficiencies for electron and muons
are 99.7\% and 99.9\% respectively, where the mis-identification rates are smaller than 0.07\%. Lepton isolation is also applied, which requires $E_{\mathrm{cone}}^{2}<4E_{\ell} + 12.2$ GeV, where $E_{\ell}$ is the energy of the lepton, and $E_{\mathrm{cone}}$ is the energy  within a cone  of $\cos~\theta_{\mathrm{cone}}>0.98$ around the lepton. The jet reconstruction and flavor tagging are done by LCFIPlus software package ~\cite{Suehara:2015ura}, which integrates vertex finding, jet reconstruction and flavor tagging. Leptons are removed from the ArborPFO before the jet reconstruction. Jets are reconstructed by Durham algorithm ~\cite{Catani:1991hj}. Exclusive clustering is performed to reconstruct exactly four jets in the final state.

\section{Event selection }
\label{sec:selection}

Events are selected targeting four $b$-jets coming from the singlet-like exotics Higgs in association with two same flavor opposite-sign leptons coming from the $Z$-boson decay. In the beginning, some loose pre-selections are applied followed by a more sophisticated multivariate approach using a Boosted Decision Tree (BDT), trained to classify signal and background events. Both electron and muon channels are considered in this analysis. Each event is required to have two well-isolated leptons with opposite charges and each having energy higher than 20 GeV. The polar angle between the two leptons is required to be within the range of $|\cos~\theta_{e^{+}e^{-}}|<0.71$ or $|\cos~\theta_{\mu^{+}\mu^{-}}|<0.81$. The angle between two isolated tracks is required to satisfy $\cos~\phi>-0.74$ for electron and muon channel respectively. The angular cuts reduce the contributions coming from lepton pair production. The invariant mass of the lepton system is required to be within the $Z$-mass window of 77.5 - 104.5 GeV. To further suppress the non-Higgs backgrounds, the recoil mass of the two lepton system is required to be within the range of $M_{\mathrm{recoil}}^{\ell \bar{\ell}}\in[124,140]$ GeV, where the recoil mass is defined in Eq. \ref{equ:recoil mass}.

\begin{equation}
  M_{\mathrm{recoil}}^{\ell\bar{\ell}}=\sqrt{(\sqrt{s}-E_\ell-E_{\bar{\ell}})^2-(\Vec{P}_\ell+\Vec{P}_{\bar{\ell}})\cdot(\Vec{P}_\ell+\Vec{P}_{\bar{\ell}})} \label{equ:recoil mass}
\end{equation}

Followed by these basic selections, two different approaches are taken to apply further selection: a traditional cut-based approach and a multivariate approach using the Boosted Decision Tree (BDT). First the cut-based selections are described followed by the BDT description.

The particles left in the event are used to reconstruct exactly four jets with polar angle $\theta_{\mathrm{jet}}$ within the range of $|\cos ~\theta_{\mathrm{jet}}|< 0.96$. The four reconstructed jets are required to contain at least 40 particles with each of these particles having an energy no less than 0.4 GeV to suppress fake jets. Selected jets are further required to pass some energy threshold: $E_{j_{1}} > 32$ GeV, $20 < E_{j_{2}} < 55$ GeV, $E_{j_{3}} < 46 $ GeV, $E_{j_{4}} < 39$ GeV. Flavor tagging of these jets is achieved by the multi-variable-based flavor tagging toolkit in LCFIPLUS. Training for the $b$-tagging algorithm is implemented with the gradient boosted decision trees (GBDT) method using  jet kinematic variables, impact parameters of tracks and secondary vertex parameters. The $b$-tagging models returns a $b$-likeness for each jet. The values are between 0 and 1, where a higher value indicates the jet is more $b$-like. The $b$-likeness of the individual jets ($L_{bi}$) are used to construct a combined $b$-likeness, defined as

\begin{equation}
    f_{b}=\frac{L_{b1}L_{b2}L_{b3}L_{b4}}{L_{b1}L_{b2}L_{b3}L_{b4}+\left(1-L_{b1}\right)\left(1-L_{b2}\right)\left(1-L_{b3}\right)
    \left(1-L_{b4}\right)}. \label{equ:blikeness}
\end{equation}

\begin{figure}[!htb]
    \centering
    \includegraphics[width=0.5\textwidth]{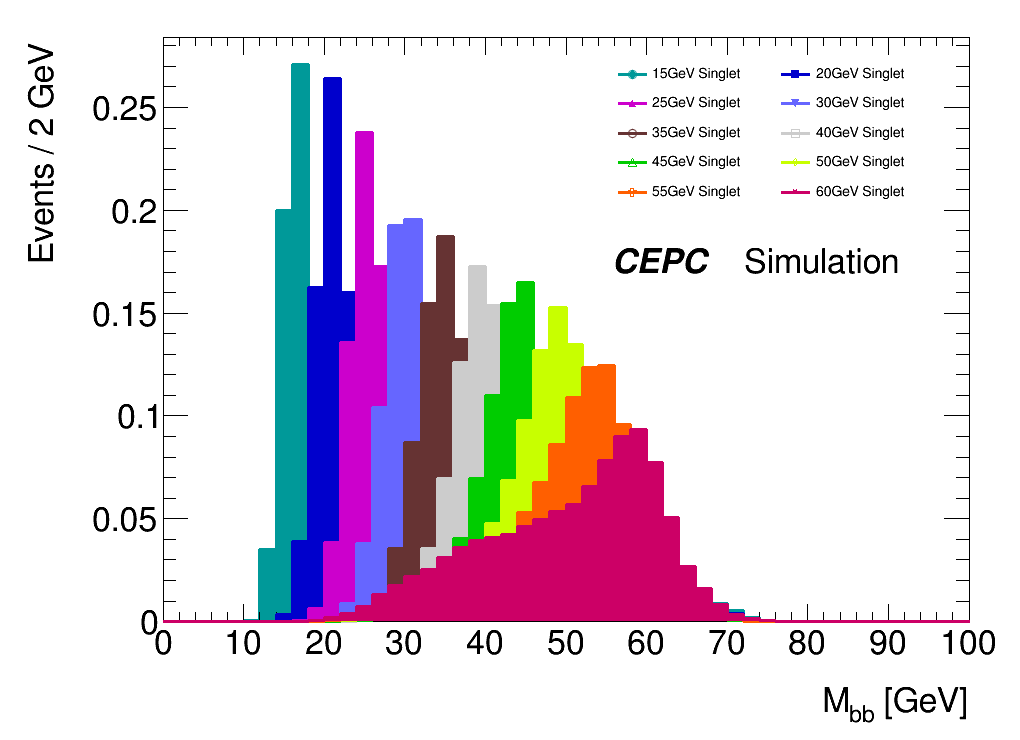}
    \includegraphics[width=0.49\textwidth]{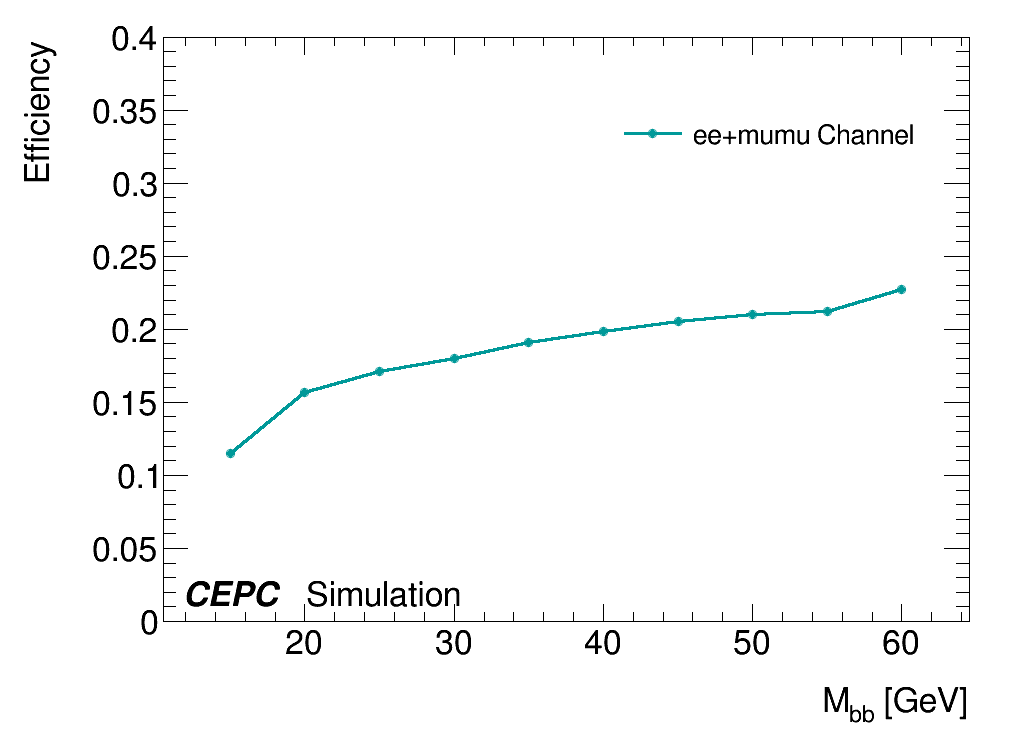}
    \caption{Normalized singlet-like exotics Higgs mass distributions ($m_{s}$) for reconstructed $H \to ss \to 4b$ events for $m_{s}$ between 15 GeV and 60 GeV (left). Signal selection efficiencies as a function of $m_{s}$ (right). }
    \label{fig:sigmass_eff}
\end{figure}

To reject the non $b$-jet backgrounds more efficiently, a $b$-jet inefficiency is defined as $\left(1 - f_{b}\right)$. Selected events are required to have $\log (1 - f_{b}) < -4.0$. Selected events are used to reconstruct the singlet scalar. All possible combination of jet pairs are check and the combination of jet pairs with lowest mass difference is chosen. Reconstructed signal mass distributions are shown in Fig. \ref{fig:sigmass_eff}. The number of events surviving after each selection criteria in the cut-based analysis is shown in Table \ref{tab:cutflow}.
 
In the other approach, BDTs are trained to separate the singlet scalar Higgs decay signals from the backgrounds of SM processes. Apart from all the variables mentioned in the cut-based approach, some additional variables with discriminating power such as the opening angles between the two reconstructed exotic Higgs and the minimum distance between the particles used in the jet reconstruction are used for BDT training. Separate BDTs are trained for different signal mass. In total there are ten BDTs. The training set consists of 30,000 signal events and 40,000 background events. Later the BDT score is used for doing the statistical analysis. 

\begin{table}[!htb]
    \centering
    \caption{Cut-flow table showing the number of events from both electron and muon channels after each selection for all the major backgrounds and signal with $m_{s} =  30$ GeV.
    \label{tab:cutflow}}
    \vspace{0.3cm}
    \begin{tabular}{l|c|c|c|c}
    \toprule
    Selection & Signal ($m_{s} = 30$ GeV) & $\ell \ell Hbb$ & other $\ell \ell H$  & non Higgs\\
    \midrule
    Original & 8865 & $2.92\times10^4$ & $2.41\times10^4$ & $3.79\times10^7$ \\
    Lepton pair selection & 6042 & $1.83\times10^4$ & $1.20\times10^4$ & $1.32\times10^6$ \\
    Lepton pair mass & 5537 & $1.65\times10^4$ & $1.07\times10^4$ & $6.17\times10^5$ \\
    Jet selection and pairing & 4054 & 7947 & 4661 & 3698 \\
    B-inefficiency & 2210 & 131 & 15 & 14 \\
    \bottomrule
    \end{tabular}
\end{table}

The background contributions are divided into three categories: Higgs processes with $bb$ final state ($\ell \ell Hbb$), other Higgs processes ($\ell \ell H$), and non-Higgs processes. Since $H\to bb$ process alone forms a large background, it is considered as a separate group; all other Higgs decay modes are grouped into a different category.

\section{Uncertainty estimation}
\label{sec:uncertainty}

The systematic uncertainties from luminosity ~\cite{Smiljanic:2021bbb} and lepton identification~\cite{Yu:2017mpx} are considered to be small.  So these uncertainties are ignored in this analysis. Systematic uncertainties coming from fixed parameters, including the dominant $\ell \ell Hbb$ background, other Higgs processes ($ \ell \ell H$), and other non-Higgs SM processes are considered. A conservative approach is taken to estimate these uncertainties. The event yields of the dominant $\ell \ell Hbb$ background process are varied up and down by 5\%, whereas the other background processes are varied by 100\% to estimate this systematic uncertainty. Systematic uncertainty coming from the flavor tagging is considered in this analysis. It is estimated based on the recipe described in Ref.~\cite{Bai_2020}. This study was done on the $ZZ\rightarrow q\bar{q}+\mu^+\mu^-$ control sample and the estimated uncertainty was 0.78\%. Since the final state in this study contains more jets, slightly higher conservative  1\% flat uncertainty is applied as the flavor tagging uncertainty. Jet energy resolution uncertainty plays an important role in this analysis. The resolution estimates from the CEPC design report are extrapolated to the lower energy region. Corresponding energy resolutions are used to smear the jet energy with a Gaussian function.

\section{Results }
\label{sec:results}

The selected events are used to set limit on the production cross-section times branching ratio of the $H \to ss \to 4b$. The selected events from both electron and muon channels are combined. The expected yields for the different SM processes are given in Table \ref{tab:cut_postfit}.

\begin{table}[!htb]
\centering
    \caption{The expected yields with total uncertainties for the various SM background processes and signal at $m_s = 30$ GeV. Events from electron and muon channels are combined.}
    \label{tab:cut_postfit}
    \vspace{0.3cm}
    \begin{tabular}{l| l}
    \toprule
    Process & Yields \\
    \midrule
    $\ell \ell Hbb$   & $130.7 \pm 104.4$ \\
    $\ell \ell H$  & $15.1 \pm 19.5$ \\
    Non Higgs   & $14.3 \pm 18.6$ \\
    \midrule
    Total Bkg & $160.1 \pm 107.8$ \\
    \midrule    
    Signal ($m_{s} = 30$ GeV) & $2210.1 \pm 1731.1$ \\
    \bottomrule
    \end{tabular}
\end{table}

\begin{figure}[hbt]
\centering
    \includegraphics[width=0.48\textwidth]{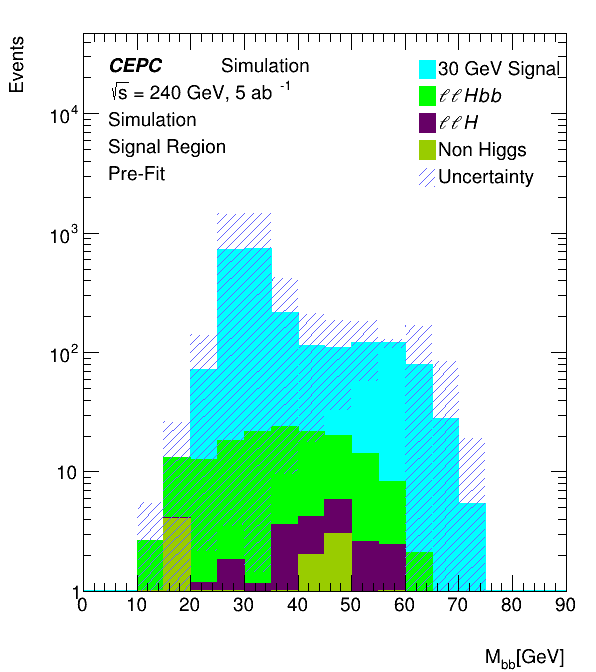}
    \includegraphics[width=0.48\textwidth]{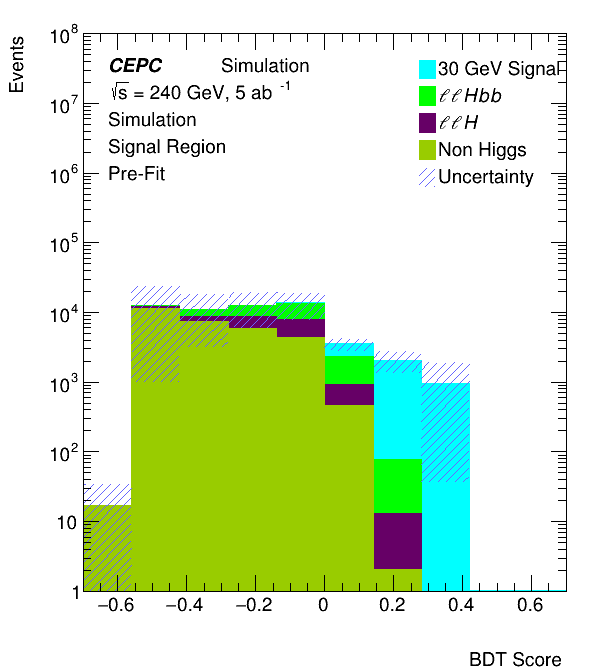}
    \caption{Distribution of reconstructed singlet-Higgs mass (left) and BDT score (right) for the combined electron and muon channel. the uncertainty band includes both statistical uncertainty and systematic uncertainties. The signal distributions for $m_{s}$ = 30 GeV are shown in cyan histograms.}
    \label{fig:discriminants_prefit}
\end{figure}

The output of the BDT classifier is used as the discriminant as described in section \ref{sec:selection}. Whereas for the cut-based analysis $m_{s}$ is used. The distributions of $m_{s}$ and the BDT score are shown on Fig. \ref{fig:discriminants_prefit}.

\begin{figure}[!htb]
\centering
    \includegraphics[scale=0.7 ]{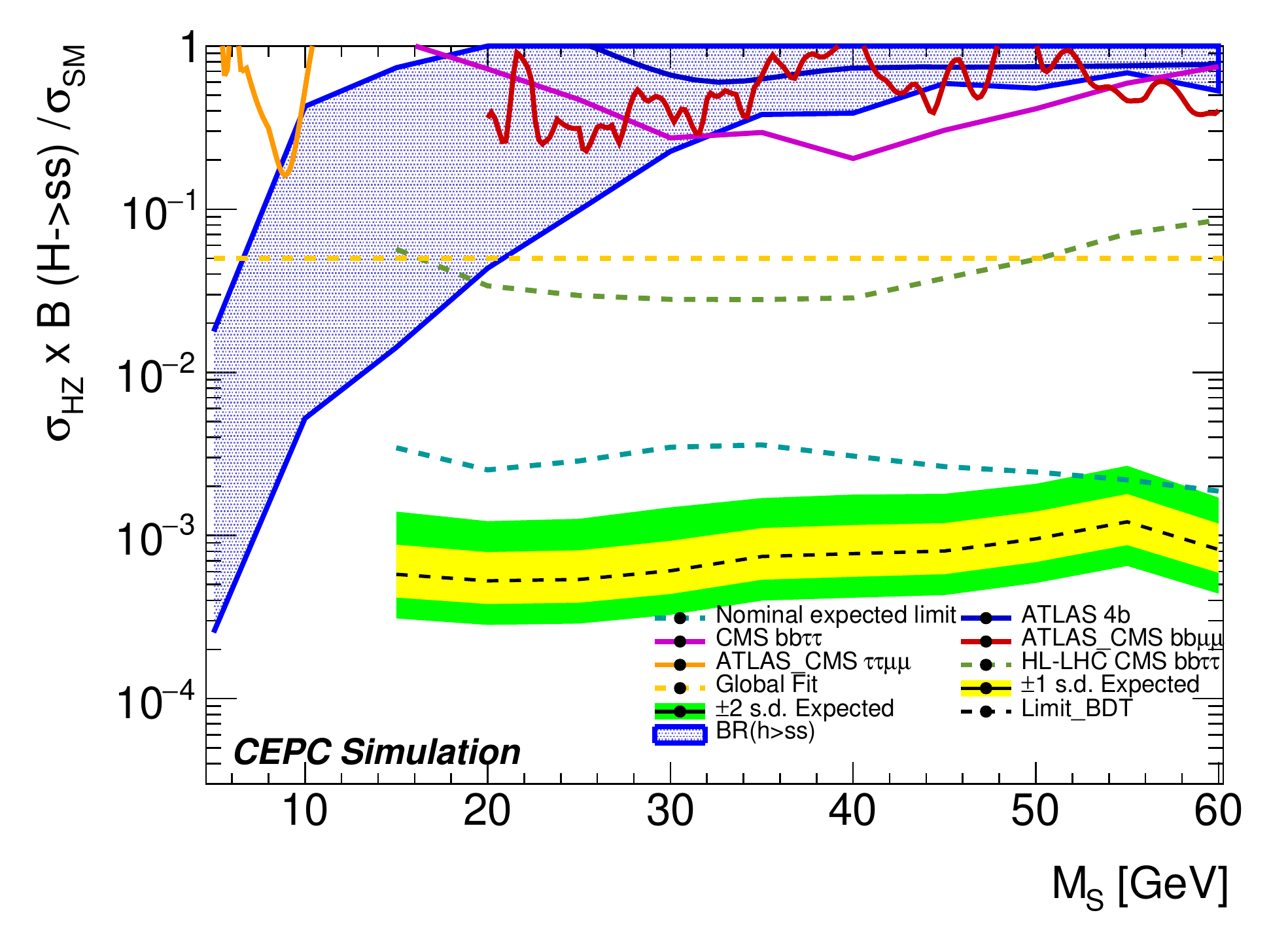}
    \caption{Expected upper limits on $\sigma_{ZH}\times \mathrm{B} (H \to ss \to 4b)$ as a function of singlet mass. Limits from the BDT analysis are shown in blue dotted line. Limits obtained from the cut based analysis is also shown for reference in cyan dotted line. The blue shaded region shows points predicting a strong first-order EWPT with successful tunneling obtained from numerical scans~\cite{kosaczuk_2020}. Leading current (solid)
    and projected (dashed) LHC sensitivities are shown in the $\tau\tau\mu\mu$ (orange) ~\cite{ATLAS:2015unc, CMS:2019spf} , $bb\mu\mu$ (pink) ~\cite{ATLAS:2018emt, CMS:2018nsh}, $bb\tau\tau$ (magenta) ~\cite{CMS-PAS-FTR-18-035, CMS:2018zvv} and $4b$ (blue) ~\cite{ATLAS:2018pvw} channels. For $bb\mu\mu$ and $\tau\tau\mu\mu$ the best limit between ATLAS and CMS is shown at any given mass point. The HL-LHC projection for the indirect limit on the total exotic branching fraction \cite{Cepeda:2019klc} is shown with the orange dashed line.
    }
    \label{fig:limit}
\end{figure}

A binned profile-likelihood fit is performed to set an expected upper limit on the $\sigma$at $95\%$ CL. All the systematic uncertainties discussed in section~\ref{sec:uncertainty} and the statistical uncertainty are considered in the fit. The systematic uncertainties are included to the likelihood using nuisance parameters. Gaussian, log-normal or Poisson priors are used in the likelihood to include the nuisance parameters.  Fig.~\ref{fig:limit} shows the upper limits on the production cross-section times branching-ratio of the singlet Higgs as a function of $m_{s}$. The upper limits, and the corresponding statistical and systematic uncertainties for the singlet signal for each $m_{s}$ are shown in Table~\ref{tab:limit}. 

\begin{table}[!htb]
\centering
    \caption{Expected upper limit on the signal cross-section times branching ratio at $95\%$ CL from both cut-based and BDT approach for combined electron and muon channel. The statistical and systematic uncertainties are shown in the form of $(\text{limit})^{+\text{stat}+\text{syst}}_{-\text{stat}-\text{syst}}$.
    \label{tab:limit}}
    \vspace{0.3cm}
    \begin{tabular}{c|c|c}
    \toprule
    \multicolumn{3}{c}{ Expected upper limits on $\sigma_{ZH}\times \mathcal{B}(H\rightarrow ss)$ }\\
    \midrule
    \multicolumn{3}{c}{$e^+e^-H$ and $\mu^+\mu^-H$ combined}\\
    \midrule
    $m_{s}$ [GeV] & Cut-based & BDT\\
    \midrule
    15 & $0.0034^{+0.0010+0.0009}_{-0.0007-0.0007}$ & $0.0006^{+0.0003+0.00007}_{-0.0002-0.00002}$ \\[4pt]
    20 & $0.0025^{+0.0008+0.0008}_{-0.0005-0.0005}$ & $0.0005^{+0.0003+0.00004}_{-0.0001-0.00002}$ \\[4pt]
    25 & $0.019^{+0.0008+0.0009}_{-0.0005-0.0006}$ & $0.0005^{+0.0003+0.00004}_{-0.0001-0.00002}$ \\[4pt]
    30 & $0.0029^{+0.0008+0.0010}_{-0.0005-0.0008}$ & $0.0006^{+0.0003+0.0001}_{-0.0002-0.00004}$ \\[4pt]
    35 & $0.0034^{+0.0008+0.0009}_{-0.0006-0.0008}$ & $0.0007^{+0.0003+0.0001}_{-0.0002-0.00005}$ \\[4pt]
    40 & $0.0031^{+0.0008+0.0007}_{-0.0005-0.0007}$ & $0.0008^{+0.0004+0.0001}_{-0.0002-0.00005}$ \\[4pt]
    45 & $0.0026^{+0.0007+0.0006}_{-0.0005-0.0006}$ & $0.0008^{+0.0004+0.0001}_{-0.0002-0.00005}$\\[4pt]
    50 & $0.0024^{+0.0007+0.0004}_{-0.0005-0.0005}$ & $0.0009^{+0.0004+0.0002}_{-0.0003-0.00008}$\\[4pt]
    55 & $0.0022^{+0.0006+0.0003}_{-0.0004-0.0004}$ & $0.0012^{+0.0005+0.0003}_{-0.0003-0.00015}$\\[4pt]
    60 & $0.0019^{+0.0006+0.0002}_{-0.0004-0.0003}$ & $0.0008^{+0.0004+0.00003}_{-0.0002-0.000005}$\\[4pt]
    \bottomrule
    \end{tabular}
\end{table}

\section{Conclusion}
\label{sec:conclusion}

This paper presents a search for exotic decays of the Higgs boson into a pair of spin-zero singlet-like particles, $H \to s s$, where the $s$-boson decays into two $b$-quarks. The search focuses on processes in which the Higgs boson is produced in association with a $Z$ boson that decays leptonically. The study is done in a scenario of analysing 5000 fb$^{-1}$  $e^+ e^-$ collision data at $\sqrt{s} = 240 $ GeV in CEPC. The analysis uses several kinematic variables for both combined in a multivariate BDT  discriminant in the signal region. A cut-based approach is also studied for comparison. Upper limits at 95\% CL are derived for the $ZH$ production cross-section times branching ratio of the decay $H \to s s \to 4 b$. The combined expected upper limit for promptly decaying $m_{s}$-bosons ranges from 0.041 for $m_{s}$ = 15 GeV to 0.0016 for $m_{s}$ = 60 GeV. This realistic study yields a weaker exclusion limit compared to the projected values in Ref.~\cite{kosaczuk_2020}. Hence, it is crucial to carry out the study with a dedicated simulation using realistic detector characteristics and experimental uncertainties. The study with $4b$ final state at the CEPC could conclusively test the possibility of an SFOEWPT in the extended-SM with a light singlet of mass as low as 20 GeV. This is a significantly greater reach than can be achieved at the LHC  at that low mass.

\acknowledgments

We are grateful to Yu Bai, Gang Li and Manqi Ruan for several useful conversations during the course of this work.
Michael Ramsey-Musolf and Yanda Wu were supported in part under National Natural Science Foundation of China grant no. 19Z103010239.

\bibliography{Reference}

\end{document}